\documentclass[letter]{aa}
\usepackage[varg]{txfonts}

\usepackage{graphicx}
\usepackage{natbib}
 \bibpunct{(}{)}{;}{a}{}{,} 
 \usepackage{hyperref}
\hypersetup{
    colorlinks,
    linkcolor={black},
    citecolor={black},
    urlcolor={blue}
}

\begin{document}
\title{Alpha particle thermodynamics in the inner heliosphere fast solar wind}

\author{
D. Stansby\inst{\ref{inst:imperial}} \and
D. Perrone\inst{\ref{inst:imperial}} \and
L. Matteini\inst{\ref{inst:meudon}} \and
T. S. Horbury\inst{\ref{inst:imperial}} \and
C. S. Salem\inst{\ref{inst:SSL}}
}
\institute{
Department of Physics, Imperial College London, London, SW7 2AZ, United Kingdom \\ \email{david.stansby14@imperial.ac.uk} \label{inst:imperial} \and
LESIA, Observatoire de Paris, Universit\'e PSL, CNRS, Sorbonne Universit\'e, Univ. Paris Diderot, Sorbonne Paris Cit\'e, 5 place Jules Janssen, 92195 Meudon, France \label{inst:meudon} \and
Space Sciences Laboratory, University of California, Berkeley, CA 94720, USA \label{inst:SSL}
}

\abstract
{Plasma processes occurring in the corona and solar wind can be probed by studying the thermodynamic properties of different ion species. However, most in-situ observations of positive ions in the solar wind are taken at 1~AU, where information on their solar source properties may have been irreversibly erased.}
{In this study we aimed to use the properties of alpha particles at heliocentric distances between 0.3 and 1 AU to study plasma processes occurring at the points of observation, and to infer processes occurring inside 0.3 AU by comparing our results to previous remote sensing observations of the plasma closer to the Sun.}
{We reprocessed the original Helios positive ion distribution functions, isolated the alpha particle population, and computed the alpha particle number density, velocity, and magnetic field perpendicular and parallel temperatures. We then investigated the radial variation of alpha particle temperatures in fast solar wind observed between 0.3 and 1 AU.}
{Between 0.3 and 1 AU alpha particles are heated in the magnetic field perpendicular direction, and cooled in the magnetic field parallel direction. Alpha particle evolution is bounded by the alpha firehose instability threshold, which provides one possible mechanism to explain the observed parallel cooling and perpendicular heating. Closer to the Sun our observations suggest that the alpha particles undergo heating in the perpendicular direction, whilst the large magnetic field parallel temperatures observed at 0.3~AU may be due to the combined effect of double adiabatic expansion and alpha particle deceleration inside 0.3 AU.}
{}

\maketitle

\section{Introduction}
The solar wind is a tenuous magnetised plasma flowing from the surface of the Sun to the edge of the heliosphere. Although protons form the bulk of the positive ions in the fast solar wind by number density ($\sim$~95\%), alpha particles (doubly ionised helium) comprise $\sim$ 5\% by ion number density, and therefore $\sim$ 10\% by ion charge density, and $\sim$~20\% by mass density \citep{Robbins1970, Kasper2007a}. In the weakly collisional fast solar wind at 1 AU the alphas are hotter than protons by a factor of around 5 \citep{Tracy2015, Kasper2017a}. This is often assumed to be a remnant of a process in the corona that heats heavy ions to higher temperatures than protons, for which several proposed mechanisms exist \citep[e.g.~see the review of][]{Cranmer2018}. Remote sensing observations of multiple heavy ions close to the Sun support this hypothesis \citep{Landi2009, Hahn2013}.

In addition to preferential heavy ion heating, the mechanism that heats protons in coronal holes deposits energy preferentially in the direction perpendicular to the magnetic field. This is inferred from both in-situ observations in the inner heliosphere down to 0.3 AU \citep{Marsch1982b, Matteini2007}, and remote sensing observations of neutral hydrogen (which reflect the ionised hydrogen distribution function due to frequent collisionless charge exchange) closer to the Sun \citep{Kohl1998, Dolei2016}. However, measurements of the preferential perpendicular heating of other ion species are sparse. From remote sensing, the only other observations are of oxygen ions, which are also observed to undergo preferential perpendicular heating \citep{Kohl1998, Cranmer2008}.

From in-situ measurements, 3D velocity distribution functions (needed un-ambiguously to determine both parallel and perpendicular temperatures) are typically only measured for protons and alpha particles. At 1 AU alpha particles are close to isotropy in all speed ranges \citep{Marsch1982c, Maruca2012d}, which leaves open the possibility that collisional processes and wave-particle interactions have irreversibly erased information on source properties such as the presence or absence of temperature anisotropies closer to the Sun. This means observations are needed closer to the Sun to distinguish between properties set by processes occurring in the corona and properties resulting from evolution as the solar wind propagates radially away from the Sun.

The \emph{Helios} mission, comprising two spacecraft in orbit around the Sun from 0.29 -- 1 AU, measured 3D alpha particle distribution functions, but studies using this data closer to the Sun than 0.5 AU have been limited to a handful of papers \citep{Marsch1982c, Schwartz1983a, Marsch1983, Thieme1989a, Geranios1989}. In this work the 3D alpha distribution functions measured by \emph{Helios} in the fast solar wind at 0.29 -- 0.8~AU are used to present, for the first time, a detailed study of the thermodynamics of alpha particles in the fast solar wind inside 1~AU.

\section{Data processing}
\label{sec:data}
In order to extract parameters of the alpha particle population we used the positive ion velocity distribution functions measured on board \emph{Helios} by the E1 electrostatic analyser experiment \citep{Rosenbauer2018}.

Because alpha particles have a mass to charge ratio twice that of protons, at the same velocity they have a higher energy per charge, and therefore it is possible to separate the distribution using an energy per charge threshold, above which measurements are dominated by protons and below which they are dominated by alpha particles \citep{Marsch1982c, Marsch1982b}. However, there are still a significant number of alpha particles present just below this threshold. This means taking moments of the truncated alpha particle distribution significantly underestimates the true number density and temperatures, but as long as enough of the distribution function is present, fitting avoids this problem. This is nicely illustrated by fig.~16 of \cite{Marsch1982b} where core parallel temperatures (calculated through fitting) are systematically higher than total parallel temperatures (calculated from moments). We therefore chose to fit analytic bi-Maxwellian distribution functions to the separated alpha distribution function. Note that as a result of this choice caution should be applied when comparing our results to other alpha particle results using \emph{Helios} data (including \citealt{Marsch1982c} and \citealt{Schwartz1983a}), which for the most part used numerical moments of the truncated distribution function.

We stress that, although a secondary alpha beam is sometimes observed in the solar wind \citep{Feldman1973, Feldman1993}, the limited energy resolution of the \emph{Helios} ion instrument means we do not think the fitting process was capable of distinguishing between an `alpha core' and `alpha beam' population. A subset of the fits from each individual interval were inspected by eye, and found accurately to characterise the width of the distribution (and therefore also the thermal energy) in both perpendicular and parallel to the magnetic field.

For specific details of the alpha fitting process and figures comparing the fits to the measured distribution functions see appendix \ref{app:fitting}. The fitting process returned the alpha number density ($n_{\alpha}$), velocity ($\mathbf{v}_{\alpha}$) and temperatures parallel ($T_{\alpha \parallel}$) and perpendicular ($T_{\alpha \perp}$) to the local magnetic field ($\mathbf{B}$) for each distribution function. The original distribution functions that were fit are openly available, and the code used to fit the distributions is also available and documented, making the resulting fits reproducible. The outputted fit parameters have also been made freely available for use in further research. For more details and links to both the code and data see appendix \ref{app:fitting}.

The observation dates of the 11 high speed streams used in this study are available at \url{https://doi.org/10.5281/zenodo.2359200}. To check the validity of our fitted parameters we performed the following checks against previously published observations of alpha particles in the inner heliosphere:
\begin{itemize}
	\item Thermal speeds are the same order of magnitude as those measured by the MESSENGER mission in fast solar wind \citep{Gershman2012a}, providing validation from independent instrumentation and processing.
	\item Temperature anisotropies are the same sense and order of magnitude as the fits presented in fig.~16 of \cite{Marsch1982b}, providing validation from an independent processing of the \emph{Helios} data.
	\item The alpha particle abundance (ratio of alpha number density to proton core number density) has a median of 4.3\%, which agrees well with alpha abundances measured in fast solar wind at 1 AU \citep{Kasper2007a} and beyond \citep{Liu1995}.
\end{itemize}

In the rest of this paper alpha particle properties are compared to concurrent proton properties taken from the dataset of \cite{Stansby2018h}. Note that these parameters are specific to the proton core and not the entire proton population, which means that $T_{p\perp}$ values  are systematically larger and $T_{p\parallel}$ values systematically smaller than previous estimates from \emph{Helios} data \cite[for more details see][section 4.3]{Stansby2018h}.

Proton parameters are denoted with a $p$ subscript, and alpha parameters with an $\alpha$ subscript. For each species $i= p,\alpha$, the parallel beta was calculated as $\beta_{i \parallel} = n_{i} k_{B} T_{i\parallel} / (\left | \mathbf{B} \right |^{2} / 2\mu_{0})$.

\section{Results}
\label{sec:results}
\subsection{Radial trends}
\label{sec:trends}

\begin{figure}
	\includegraphics[width=\columnwidth]{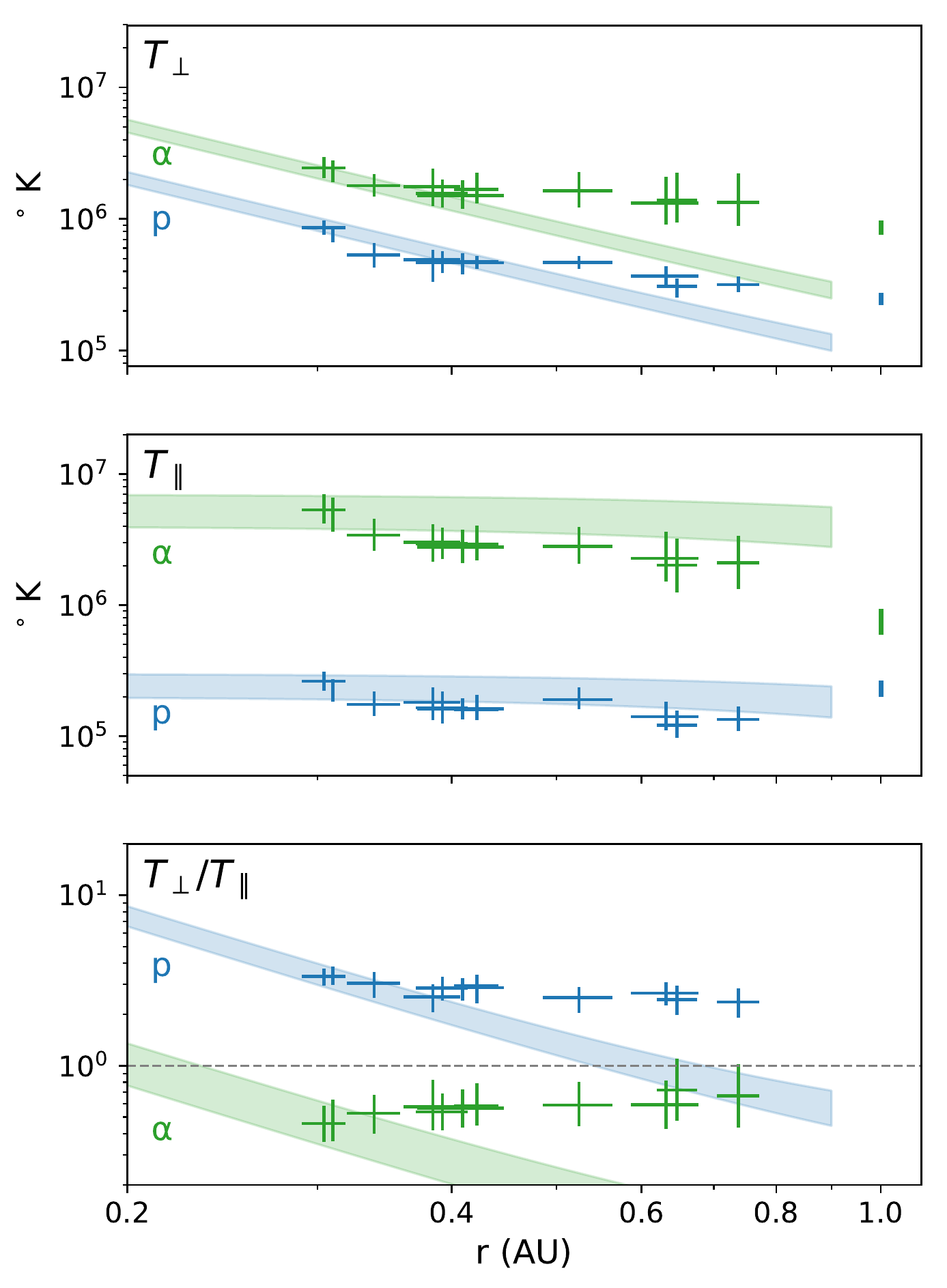}
	\caption{Radial evolution of alpha (green) and proton (blue) temperatures in fast solar wind. Each cross is measured in a single continuous high speed stream. The width of crosses gives the range of radial distances sampled in each stream. The centre of the cross is at the median value, and the lower and upper bounds are the first and third quartiles of the data. Bars at 1~AU are taken from the \emph{WIND swe\_h1}  dataset. Shaded areas are a double adiabatic prediction \citep{Chew1956} based on an ideal Parker spiral magnetic field \citep{Parker1958}, for constant solar wind speeds in the range 600~km/s -- 800 km/s.}
	\label{fig:radial}
\end{figure}

Figure \ref{fig:radial} shows the evolution of $T_{\alpha \perp}$ and $T_{\alpha \parallel}$ in the selected fast solar wind streams from 0.29 -- 0.8 AU. At 0.3 AU the alpha particles are hotter than protons in both the perpendicular direction ($T_{\alpha \perp} \approx 3 T_{p \perp}$) and the parallel direction ($T_{\alpha \parallel} \approx 18 T_{p \parallel}$).

In the absence of any collisions or wave-particle interactions the plasma temperatures are expected to follow a double adiabatic profile as the plasma propagates away from the Sun \citep{Chew1956, Matteini2011}. Figure \ref{fig:radial} shows this prediction modelled forwards from the observed temperatures at 0.3 AU. For comparison, proton temperatures with corresponding adiabatic predictions are also shown in blue, along with the range of data recorded in fast solar wind ($v_{r} > 500$ km/s) at 1 AU by \emph{WIND}.

In the perpendicular direction (fig.~\ref{fig:radial}, top panel), both the alphas and protons cool more slowly than the double adiabatic prediction, but evolve with distance in a similar manner such that $T_{\alpha \perp} / T_{p \perp}=const \approx 3$. The slower decrease of the proton perpendicular temperature in fast wind with respect to the double adiabatic prediction is well known \citep{Marsch1982b, Hellinger2011}, and our results indicate that the alpha perpendicular temperature undergoes a similarly slow decrease.

In the parallel direction (fig.~\ref{fig:radial}, middle panel) $T_{\alpha \parallel}$ decreases faster than the double adiabatic prediction, meaning there must be at least one mechanism that is cooling the alpha particles in the parallel direction between 0.3 AU and 1 AU. This parallel cooling drives a decrease from $T_{\alpha \parallel} / T_{p \parallel} \approx 18$ at 0.3 AU to $T_{\alpha \parallel} / T_{p \parallel} \approx 14$ at 0.8 AU. In combination the fast decrease of $T_{\alpha \parallel}$ and slow decrease of $T_{\alpha \perp}$ means that the alpha temperature anisotropy increases with radial distance, strongly deviating from the double adiabatic prediction (fig.~\ref{fig:radial}, bottom panel).

\subsection{Instability bounded evolution}
\label{sec:instabilities}
\begin{figure}
	\resizebox{\hsize}{!}{\includegraphics{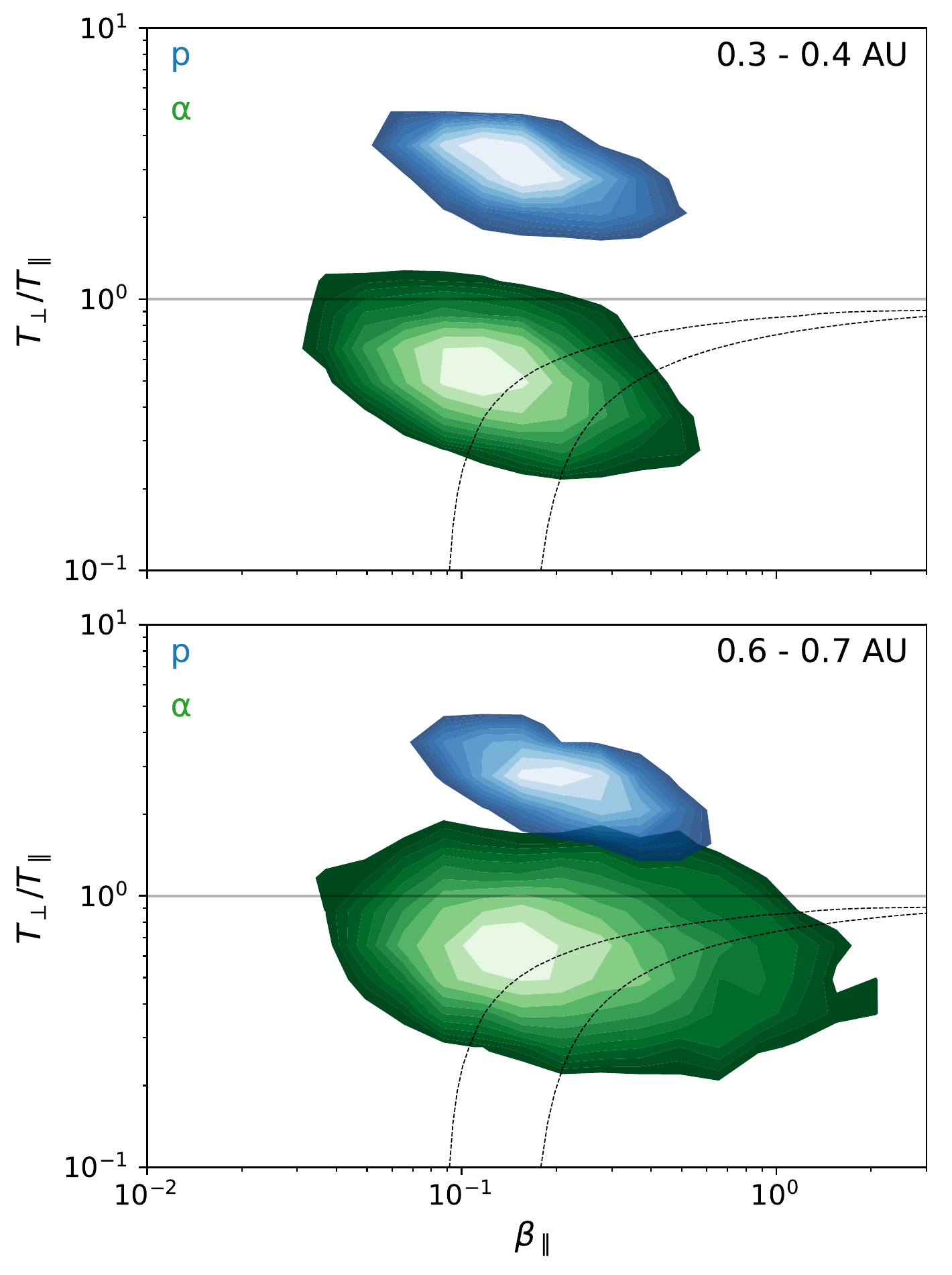}}
	\caption{Joint probability distribution of parallel beta and temperature anisotropy for protons (blue) and alphas (green) at radial distances 0.3 -- 0.4 AU (top panel) and 0.6 -- 0.7 AU (bottom panel) in fast solar wind. The contours are logarithmically spaced from the maximum bin count to 0.1 times the maximum bin count. The dashed lines show the alpha firehose instability threshold for normalised growth rates of $\gamma / \Omega_{\alpha} = 10^{-3}$ (left hand curve) and $10^{-2}$ (right hand curve) . See appendix \ref{app:instability} for details of the instability curve calculation. The large spread in alpha particle values is likely partly due to experimental uncertainties in fitting alpha number densities and parallel temperatures.}
	\label{fig:ani beta}
\end{figure}

While the protons at 0.3 AU have $T_{p \perp} / T_{p \parallel} > 1$ \citep{Marsch1982b, Marsch2004}, the alphas have $T_{\alpha \perp} / T_{\alpha \parallel} < 1$. This has previously been mentioned by \cite{Marsch1982c}, but fig.~\ref{fig:radial} shows that this feature persists across the two observed fast solar wind streams at 0.3~AU. To make the difference between proton and alpha anisotropies clearer, the top panel of fig.~\ref{fig:ani beta} shows the joint distribution of $T_{\perp} / T_{\parallel}$ and $\beta_{\parallel}$ for both protons and alpha particles in data taken at distances 0.3 AU $<$ r $<$ 0.4~AU. The alphas have similar parallel beta values to the protons ($\beta_{\alpha \parallel} \approx \beta_{p \parallel} \approx 0.1$), but much lower temperature anisotropies. 

Double adiabatic expansion of the solar wind acts to reduce $T_{\perp} / T_{\parallel}$ and increase $\beta_{\parallel}$ with radial distance \citep{Chew1956, Matteini2011}. As radial distance increases, this has the effect of moving the plasma closer to instability boundaries in the $T_{\perp} / T_{\parallel}$ -- $\beta_{\parallel}$ parameter space. At 0.6 -- 0.7 AU (fig.~\ref{fig:ani beta}, bottom panel) the protons have moved to higher $\beta_{\parallel}$ and lower anisotropies, but they only start to encounter the instability boundaries at $r \approx$ 1 AU \citep{Matteini2007, Matteini2013a}. The alphas reach an instability threshold at smaller heliocentric distances than this for two reasons: they already have low temperature anisotropies at 0.3 AU, bringing them closer to the $T_{\perp} < T_{\parallel}$ instability boundaries than protons, and the instability boundaries for alphas are located at lower beta relative to the proton instability boundaries \citep{Matteini2015a}. To confirm this, fig.~\ref{fig:ani beta} shows the numerical instability threshold for the alpha firehose instability (see appendix \ref{app:instability} for details of the threshold calculation).

The distribution of data relative to the instability thresholds at the two distances is consistent with the alpha particles being restricted from moving to lower $T_{\alpha \perp} / T_{\alpha \parallel}$ and higher $\beta_{\alpha \parallel}$ due to these properties being limited by the alpha firehose instability. This agrees with similar observations of alpha particles at 1 AU, which show a wide spread of alpha parameters bounded by theoretical instability curves \citep{Maruca2012d}. In addition, at 0.3 AU the well known anti-correlation of $T_{\perp} / T_{\parallel}$ and $\beta_{\parallel}$ present for the protons \citep{Marsch2004} is also present in the alpha particles, but by 0.6 AU the anti-correlation for the alpha particles has been removed.

\subsection{Heating rates}
\label{sec:heating}
The amount of heating or cooling relative to simple double adiabatic expansion can be roughly estimated from these observations. In order to minimise the effect of varying temperatures due to different solar wind source properties, the heating rate was evaluated using two streams observed at 0.3~AU and 0.65 AU identified by \cite{Perrone2018a} as originating from the same coronal hole (their streams A7 and A8). 

The expected temperatures due \emph{only} to adiabatic evolution depend on the ratios of number densities and magnetic field strengths at the two points of interest \citep{Chew1956}
\begin{equation}
	T_{\perp 2} = T_{\perp 1} \frac{\left | \mathbf{B}_{2} \right |}{\left | \mathbf{B}_{1} \right |}
\end{equation}
\begin{equation}
	T_{\parallel 2} = T_{\parallel 1} \left ( \frac{n_{2}}{n_{1}} \frac{\left | \mathbf{B}_{1} \right |}{\left | \mathbf{B}_{2} \right |} \right )^{2}
	\label{eq:CGL par}
\end{equation}
where the LHS are predicted temperatures and the RHS contain quantities measured at 0.3 AU (subscript `1') and at 0.65 AU (subscript `2').

The total energy density at any given distance is partitioned into the perpendicular and parallel directions: $\epsilon_{\perp} = nk_{B}T_{\perp}$, $\epsilon_{\parallel} = \frac{1}{2} nk_{B}T_{\parallel}$. These values can be evaluated at 0.65 AU using the observed number density and either the observed temperatures, or the predicted temperatures. The difference between the estimated energy density and observed energy density then gives a measure of the total energy added or removed from the plasma during transit, $\Delta \epsilon_{\perp,\parallel}$. To obtain a rough estimate for the heating rates ($Q_{\perp, \parallel}$), 
$\Delta \epsilon_{\perp, \parallel}$ was divided by the transit time $\Delta t = \Delta r / v_{\alpha r}$ where $\Delta r$ = 0.35 AU is the radial distance between the two observations and  $v_{\alpha r}$ = 725 km~s$^{-1}$ is the average alpha particle radial velocity.

Estimates of the heating rates were calculated using the median observed values of $n_{\alpha}$ and $\left | \mathbf{B} \right |$ in each stream, and therefore do not assume a speed profile or any specific magnetic field geometry in between the two points. Upper and lower limits were calculated using first and third quartile temperature values in each stream. Performing the calculations gives\footnote{values are denoted as $\left ( \textup{median}^{\textup{upper bound}}_{\textup{lower bound}} \right )$}

\begin{align}
	Q_{\perp \alpha} = + \left [ 4.11^{~10}_{~0.4} \right ] \times 10^{-17}~~\textup{Wm}^{-3} \\
	Q_{\parallel \alpha} = -\left [13.7^{~22}_{~5.7} \right ] \times 10^{-17}~~\textup{Wm}^{-3}
\end{align}
Although the range of values in each case spans about a decade, the alphas are clearly heated in the perpendicular direction and cooled in the parallel direction.

\section{Discussion}
The radial evolution of alpha particle temperatures (fig.~\ref{fig:radial}) is inconsistent with a simple double-adiabatic prediction: $T_{\alpha \parallel}$ decreases faster than predicted with radial distance and $T_{\alpha \perp}$ decreases more slowly. In addition the alpha particles lie near the thresholds for anisotropy instabilities at 0.3 AU (fig.~\ref{fig:ani beta}). The continual driving of temperature anisotropy instabilities is therefore one probable cause of the non-adiabatic evolution. These instabilities act locally to increase $T_{\alpha\perp}$ and decrease $T_{\alpha\parallel}$ according to both theory \citep[e.g.][]{Verscharen2013a, Matteini2015a, Seough2016}, and expanding box simulations of an electron-proton-alpha solar wind \citep{Hellinger2003, Hellinger2013}.

In addition to temperature anisotropy instabilities there are many other proposed mechanisms that could influence the observed non-adiabatic behaviour: alpha streaming instabilities \citep[e.g.][]{Gary2000b, Verscharen2013a, Verscharen2015, Markovskii2019}, parametric decay instabilities \citep[e.g.][]{He2016}, stochastic heating \citep[e.g.][]{Chandran2010, Wang2011}, ion cyclotron heating \citep[e.g.][]{Xie2004} or dissipation of turbulence \citep[e.g.][]{Perrone2014, Perrone2014a, Maneva2015}. Recently it has become clear that judging the stability of a plasma based solely on the distribution of data in a single anisotropy-beta parameter space is overly simplistic \citep{Klein2017}. In the future our new alpha dataset can be used with other methods to evaluate instabilities to which the alpha particles contribute \citep[e.g.][]{Bourouaine2013, Chen2016a, Klein2018}.

Previous estimates of proton heating rates in the fast solar wind from 0.3 -- 0.65 AU show evidence of heating in the perpendicular direction and cooling in the parallel direction \citep{Hellinger2011}, with
\begin{align}
	Q_{\perp p} \approx + \left [ 10^{-16} -10^{-15} \right ]~~\textup{Wm}^{-3} \\
	Q_{\parallel p} \approx -\left [10^{-16} -10^{-15} \right ]~~\textup{Wm}^{-3}
\end{align}
These proton heating/cooling estimates are an order of magnitude larger than our estimates for the alpha heating rates, however converting the heating rates to a heating rate per particle using the observed median number density ratio of $n_{\alpha} / n_{p} = 0.043$ reveals that the heating rate per particle is likely higher for alpha particles than protons. This is in agreement with solar wind simulations that show alpha particles undergoing more heating than protons \citep{Hellinger2013, Perrone2014, Valentini2016}.

\begin{figure}
	\resizebox{\hsize}{!}{\includegraphics{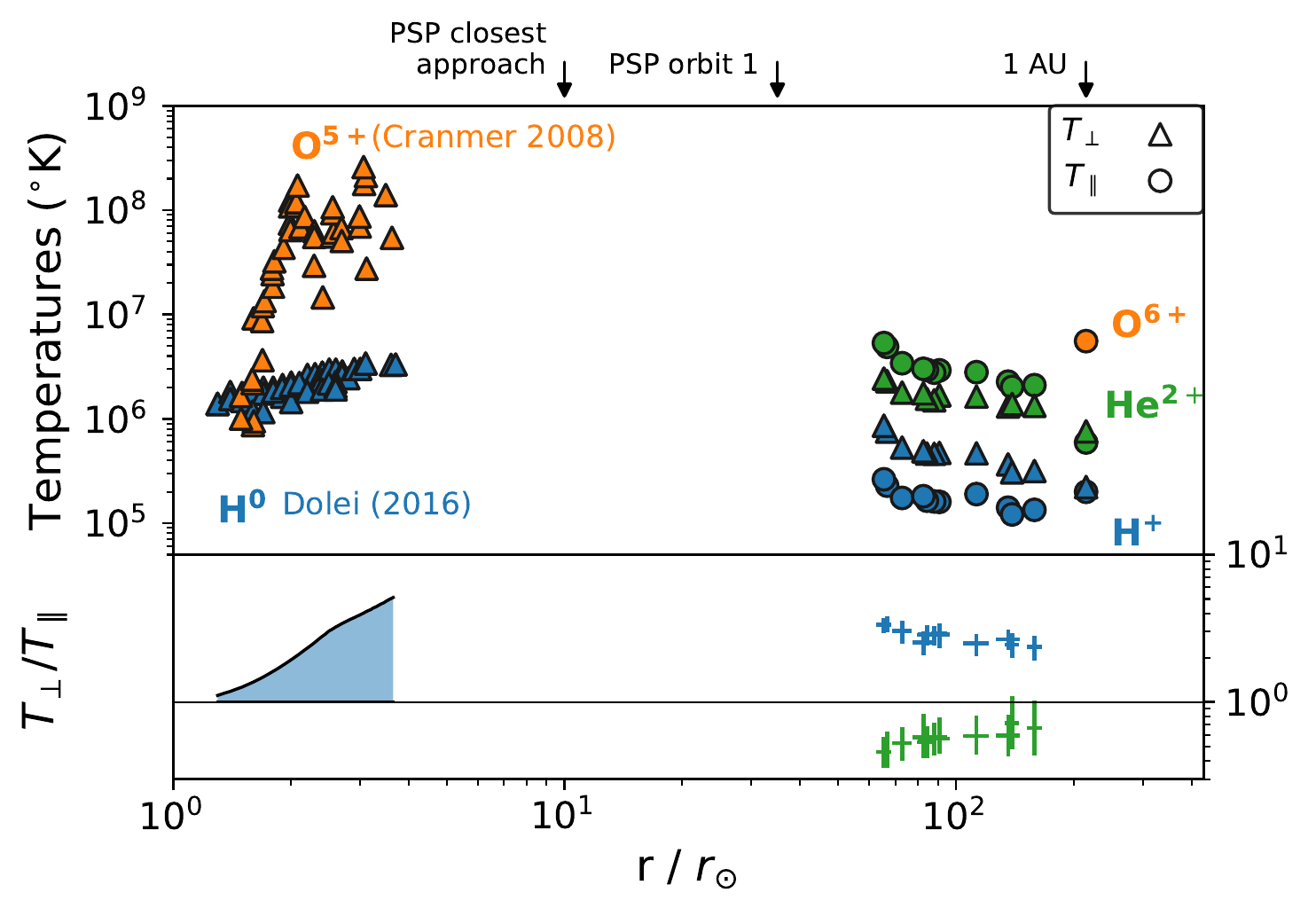}}
	\caption{Measurements of fast solar wind ion temperatures inside 1 AU. Top panel shows perpendicular (triangles) and parallel (circles) temperatures for protons (blue), alpha particles (green), and oxygen ions (orange). New results from this work are alpha particles at large distances. $T_{H^{0}\perp}$ values close to the Sun are from the polar measurements in fig.~2 of \cite{Dolei2016}. $T_{O^{5+}\perp}$ values are from fig.~5 of \cite{Cranmer2017}, originally reported by \cite{Cranmer2008}.}
	\label{fig:all r}
\end{figure}

\subsection{Evolution inside 0.3 AU}
Our results provide clues about plasma conditions closer to the Sun than 0.3 AU. Figure \ref{fig:all r} shows an overview of ion temperatures in the solar wind inside 1 AU from both in-situ and remote sensing measurements. In this work we have provided the radial variation of alpha particles from 0.3 to 1 AU, which is shown alongside remote sensing measurements of hydrogen and oxygen close to the Sun, and in situ measurements of protons from 0.3 to 1 AU and oxygen at 1 AU.

Measurements at 1 AU of weakly collisional fast solar wind indicate that heavy ion temperatures are proportional to ion mass \citep{Tracy2016a}. If we assume that this holds close to the Sun, and the alpha perpendicular temperature is less than the observed oxygen perpendicular temperature, there is an upper limit on the average rate at which $T_{\alpha \perp}$ can decrease inside 0.3 AU. This upper limit is $T_{\alpha\perp} \sim r^{-1}$, which means inside 0.3 AU the alpha particles must be heated above the double adiabatic prediction (assuming a radial magnetic field) of $\sim r^{-2}$.

Making a prediction in the parallel direction is more complicated because the double adiabatic prediction for $T_{\parallel}$ depends on evolution of both the magnetic field strength and number density (eq.~\ref{eq:CGL par}). In fast solar wind alpha particles stream faster than the protons at a large fraction of the local Alfv\'en speed. Because the Alfv\'en speed decreases with distance, the alpha particles decelerate as they travel away from the Sun \citep{Marsch1982c}. Because the radial particle flux must be conserved, $n_{\alpha} r^{2}$ must increase with distance. If the alpha particles also stream at a significant fraction of the local Alfv\'en speed inside 0.3 AU, they will also undergo deceleration between the Sun and 0.3 AU. This means that $n_{\alpha}$ would decrease slower than a 1/$r^{2}$ constant-speed spherical expansion, and as a consequence $T_{\alpha \parallel}$ would increase with distance. A CGL evolution alongside alpha particles continuously decelerating would manifest itself as a significant increase in $T_{\alpha \parallel}$, providing a mechanism to generate the high values of $T_{\alpha \parallel}$ and low values of $T_{\alpha \perp} / T_{\alpha \parallel}$ we have observed at 0.3 AU without the need to invoke additional heating or cooling mechanisms inside 0.3 AU.

Finally, although our conclusions on the evolution of alpha particle properties inside 0.3 AU are somewhat inconclusive, they provide context for measurements taken SWEAP instrument suite on Parker Solar Probe \citep{Kasper2015, Fox2015}, which will measure the properties of alpha particles inside 0.25 AU, filling in a large part of the unexplored region in fig.~\ref{fig:all r}.


\begin{acknowledgements}
The authors thank Petr Hellinger for helpful discussions, and thank the anonymous referee for comments that improved discussion of the results. D.~Stansby is supported by STFC studentship ST/N504336/1. D.~Perrone and T.~S.~Horbury are supported by STFC grant ST/N000692/1. L.~Matteini is supported by the Programme National PNST of CNRS/INSU co-funded by CNES. Work at UC Berkeley was supported by NASA grants NNX14AQ89G and 80NSSC18K0370.
\end{acknowledgements}

\bibliographystyle{aa}
\bibliography{/Users/dstansby/Dropbox/library}

\begin{appendix}
\section{Alpha particle velocity distribution function fitting}
\label{app:fitting}
\begin{figure}
	\resizebox{\hsize}{!}{\includegraphics{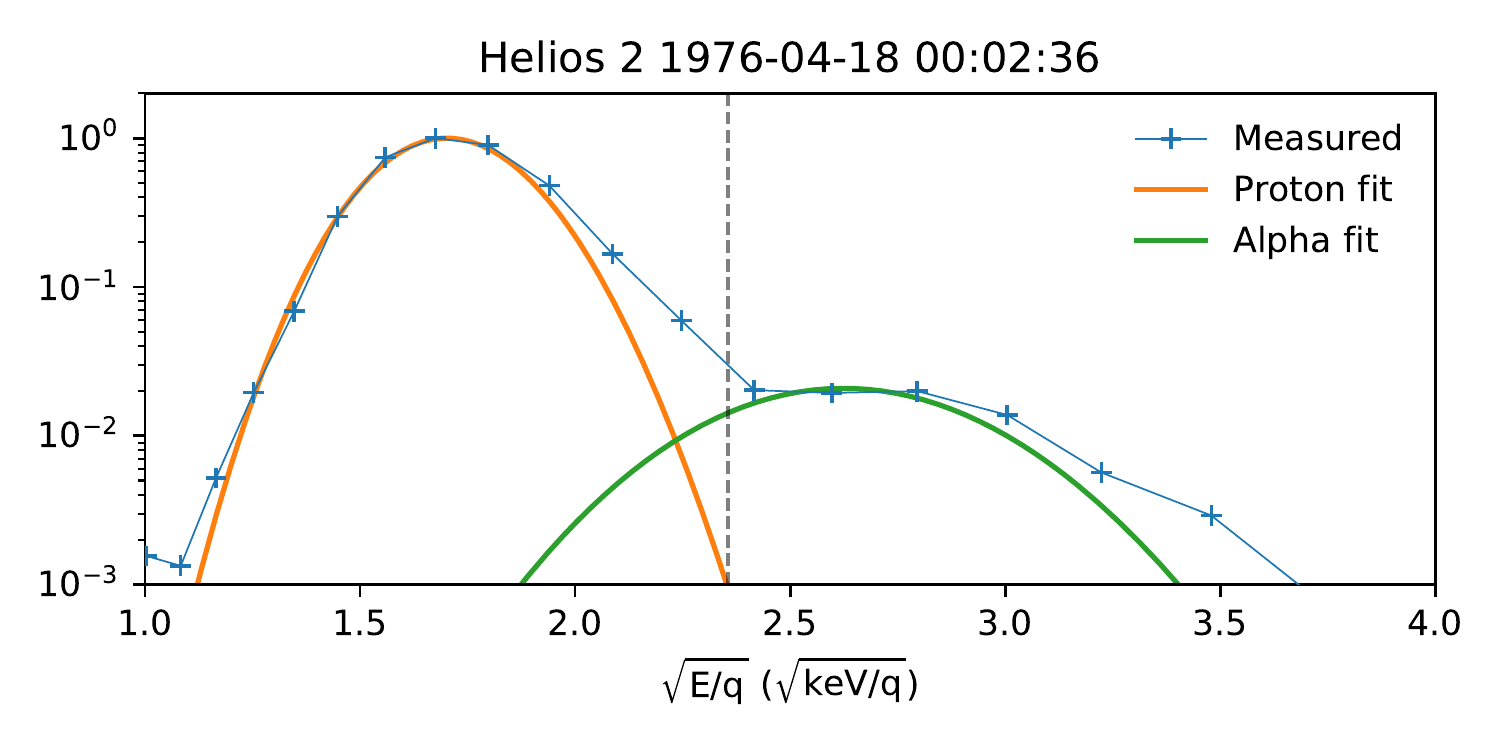}}
	\caption{An example solid angle integrated distribution function (blue line with crosses), along with the previously fitted proton distribution function (orange), and fitted alpha particle distribution function (green). The vertical dashed black line shows the threshold above which the distribution function was dominated by alpha particles. x-axis units are chosen to be linearly proportional to particle velocity for a given particle mass and charge, and the y-axis is normalised to the peak value.}
	\label{fig:1d alpha fit}
\end{figure}

To separate the alpha particle velocity distribution function (VDF) from the proton VDF, the difference in the ratio of the on board electrostatic analyser and current analyser was used \citep[see fig.~1,][]{Marsch1982c}. In this way the ratio of particle flux to charge flux was evaluated; this is twice as high for the alphas than the primary proton peak. The first energy per charge bin in the energy spectra above the proton peak where the particle to charge flux dropped below 0.8 was taken as the dividing point between the proton and alpha VDFs. If any of the angular bins at this energy were too close to the fitted proton core, the threshold was moved outward until it was at least 2 proton core thermal widths away from the proton core peak. All points with energies above (and not including) this energy per charge were assumed to be part of the alpha distribution. Figure \ref{fig:1d alpha fit} shows an example energy spectrum along with the location of the threshold.

Once the alpha part of the VDF was separated, we followed a similar process to \cite{Stansby2018h} who fitted the proton core population in the same measured VDFs. The distribution was rotated into a field aligned frame, and a bi-Maxwellian distribution function fitted (see eq.~1 of \citealt{Stansby2018h} for its functional form). The fitting process returned 6 fit parameters, from which the alpha number density ($n_{\alpha}$), velocity ($\mathbf{v}_{\alpha}$) and temperatures parallel ($T_{\alpha \parallel}$) and perpendicular ($T_{\alpha \perp}$) to the magnetic field were calculated.

\begin{figure}
	\resizebox{\hsize}{!}{\includegraphics{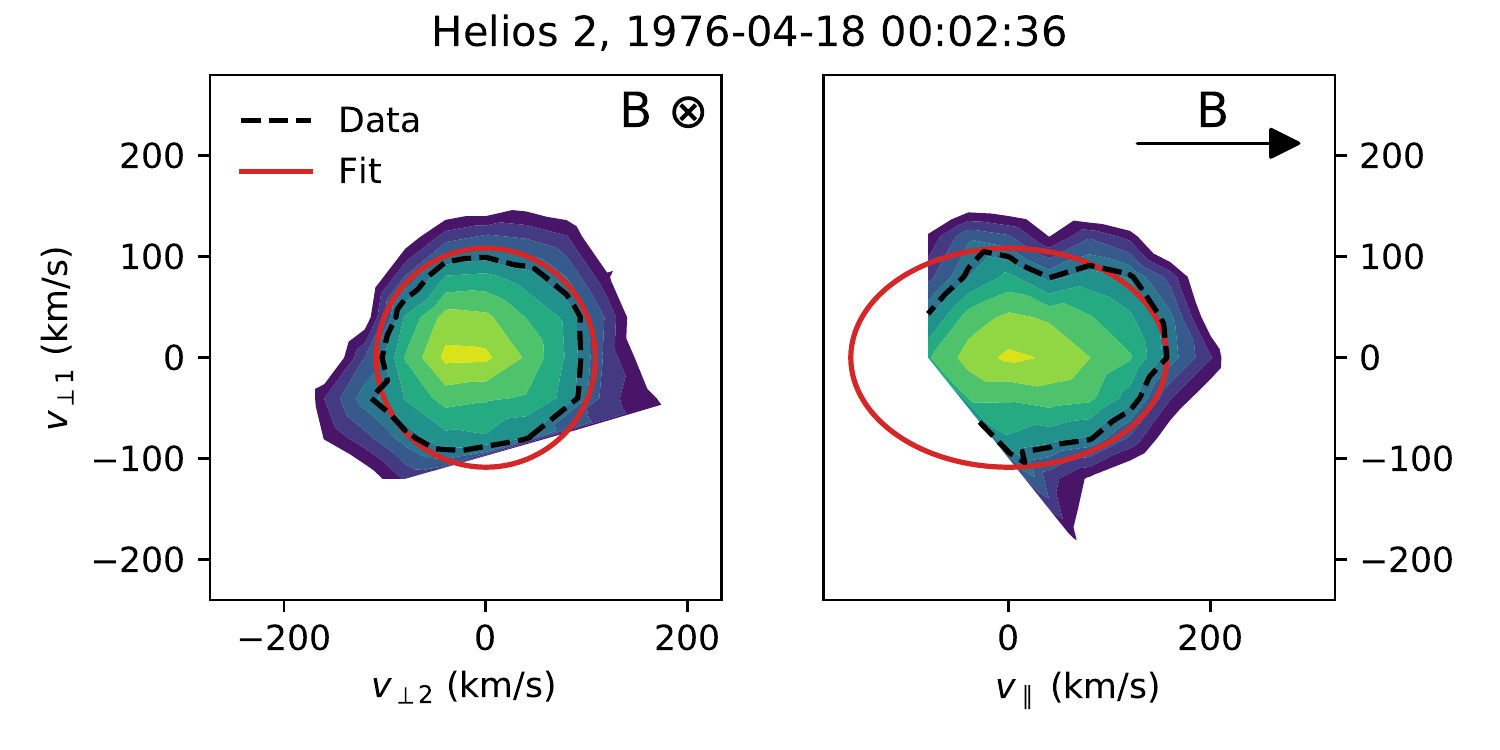}}
	\caption{Cuts of a 3D alpha velocity distribution function in the plane perpendicular to the magnetic field (LH panel) and a plane that contains the magnetic field (RH panel), centred on the fitted bulk speed. The sharp cut--offs in both plots are artificial and due to removal of the proton-dominated areas of the distribution function. The contours are logarithmically spaced. Dashed black lines show the contour at $1/e$ times the maximum value, which is one thermal width away from the bulk speed, and red ellipses show the corresponding contour for the fitted bi-Maxwellian.}
	\label{fig:alpha fit}
\end{figure}

Figure \ref{fig:1d alpha fit} shows a comparison of the solid angle integrated distribution function, along with the corresponding fitted distribution. Figure \ref{fig:alpha fit} shows two cuts of the experimentally measured distribution function in the planes perpendicular (LH panel) and parallel (RH panel) to the local magnetic field, along with a comparison between an isocontour of the data and the fitted distribution function.

The resulting values of $T_{\parallel \alpha}$ and $T_{\perp \alpha}$ are significantly larger than values reported previously \citep{Marsch1982c}. This is because previous values were calculated from numerical integration of the truncated distribution function (ie. only points to the right of the dashed black line in figure \ref{fig:1d alpha fit}). This method explicitly assumes that no alpha particles are present below the threshold, underestimating the width, and therefore the temperature, of the true alpha particle distribution.


The original distribution functions are available at the \emph{Helios} data archive\footnote{\url{ftp://apollo.ssl.berkeley.edu/pub/helios-data/E1_experiment/helios_original/}}. The code used to fit the alpha particle velocity distribution functions\footnote{\url{https://doi.org/10.5281/zenodo.2543672}} and the fitted alpha particle parameters\footnote{\url{https://doi.org/10.5281/zenodo.2358792}} are also available on the \emph{Helios} data archive. Researchers interested in using alpha particle parameters from intervals of \emph{Helios} data not analysed in this letter are encouraged to contact the first author.
\section{Instability curve calculation}
\label{app:instability}
To generate the instability curves shown in fig.~\ref{fig:ani beta} the New Hampshire Dispersion Solver \citep[NHDS,][]{Verscharen2018} was used. The 3-fluid proton-alpha-electron dispersion relation was numerically solved for a range of wavevectors, and the highest growth rate, $\gamma$, determined as a function of the input plasma parameters. The input parameters are listed in table \ref{tab:summary}, and were chosen based on average values observed in the proton and alpha data. Variations in the observed proton parameters between 0.3~AU and 0.65 AU result in very small changes to the instability threshold, so for simplicity only one threshold was calculated for both distances based on average values observed at 0.3 AU. The electron drift speed and number density were set to enforce charge neutrality and zero net current. Only instabilities with propagation parallel to the magnetic field were investigated, as for $T_{\alpha \perp} / T_{\alpha \parallel} < 1$ they have higher growth rates than the obliquely propagating modes \citep{Maruca2012d}.

The alpha temperature anisotropy and parallel beta were varied to calculate $\left [ \Omega_{c\alpha}^{-1} \gamma \right ] \left ( T_{\alpha\perp} / T_{\alpha\parallel}, \beta_{\alpha\parallel} \right )$ on a fixed grid  \citep[e.g.~as in][]{Hellinger2006a}, where $\Omega_{c\alpha}$ is the alpha particle gyrofrequency given by $\Omega_{c\alpha} = q_{\alpha} \left | \mathbf{B} \right | / m_{\alpha}$. The $\gamma = 10^{-3} \Omega_{c\alpha}$ and $\gamma = 10^{-2} \Omega_{c\alpha}$ contours were then interpolated from this grid, and are shown on fig.~\ref{fig:ani beta}.

\begin{table}
	\caption{Plasma parameters used to derive the instability curves shown in fig.~\ref{fig:ani beta}. The Alfv\'en speed is defined as $v_{A} = \left | \mathbf{B} \right | / \sqrt{\mu_{0} m_{p} n_{p}}$, and $v_{\alpha}$ is the field aligned drift of the alpha particles.}
	\label{tab:summary}
	\begin{tabular}{ll}
   		Parameter & Value \\
		\hline
		$\theta_{kB}$								& 0.0001 $^{\circ}$	\\
		$v_{A}$									& 150 km/s \\
		$v_{p} / v_{A}$								& 0 \\
		$v_{\alpha} / v_{A}$							& 0.65 \\
		$n_{\alpha} / n_{p}$							&  0.05 	\\
		$T_{p\perp} / T_{p\parallel}$					& 3			\\
		$\beta_{p\parallel}$							& 0.15		\\
		$T_{e\perp} / T_{e \parallel}$					& 1 \\
		$\beta_{e\parallel}$							& 1 
	\end{tabular}
\end{table}

\section{Data and Software}
The \emph{WIND} particle data used in this study are available at \url{ftp://spdf.gsfc.nasa.gov/pub/data/wind/swe/swe_h1/}. The \emph{Helios} distribution function data are available on the \emph{Helios} data archive (\url{http://helios-data.ssl.berkeley.edu/}), along with the corresponding fitted proton parameters (also available at \url{https://zenodo.org/record/1009506}). For details and links to alpha particle code and data see appendix \ref{app:fitting}.

Data were retrieved using HelioPy v0.6.3 \citep{Stansby2018e} and processed using astropy v3.1 \citep{Price-Whelan2018}. Figures were produced using Matplotlib v3.0.2 \citep{Hunter2007, Caswell2018}. Code to reproduce figures 1 to 3 can be found at \url{https://github.com/dstansby/publication-code}.

\end{appendix}

\end{document}